\documentclass[final]{aipproc}

\layoutstyle{6x9}

\usepackage{amsbsy}

\newcommand{\pre}{Phys. Rev. E}
\newcommand{\apj}{Astrophysical Journal}
\newcommand{\mnras}{MNRAS}
\newcommand{\prl}{Phys. Rev. Letters}

\begin{document}

\title{Aspects of Density Fluctuations in Compressible MHD Turbulence}



\author{Grzegorz Kowal}{
  address={Department of Astronomy, University of Wisconsin, Madison, WI},
  email={kowal@astro.wisc.edu}
}
\author{A. Lazarian}{
  address={Department of Astronomy, University of Wisconsin, Madison, WI},
  email={lazarian@astro.wisc.edu}
}

\begin{abstract}
We study scaling relations of compressible isothermal strongly magnetized turbulence using numerical simulations with resolution 512$^3$. We find a good correspondence of our results with the Fleck (1996) model of compressible hydrodynamic turbulence. In particular, we find that the density-weighted velocity, i.e. $\boldsymbol{u} \equiv \rho^{1/3} \boldsymbol{v}$, proposed in Kritsuk et al. (2007) obeys the Kolmogorov scaling, i.e. ${\cal E}_{u}(k)\sim k^{-5/3}$ for the high Mach number turbulence. Similarly, we find that the exponents of the third order structure functions for $\boldsymbol{u}$ stay equal to unity for all Mach numbers studied. The scaling of higher order correlations obeys the She-L\'{e}v\^{e}que (1994) scalings corresponding to the two-dimensional dissipative structures, and this result does not change with the Mach number either. In contrast to velocity $\boldsymbol{v}$ which exhibits different scaling parallel and perpendicular to the local magnetic field, the scaling of $\boldsymbol{u}$ is similar in both directions. In addition, we find that the peaks of density create a hierarchy in which both physical and column densities decrease with the scale in accordance to the Fleck (1996) predictions. This hierarchy can be related ubiquitous small ionized and neutral structures (SINS) in the interstellar gas. We believe that studies of statistics of the column density peaks can provide both consistency check for the turbulence velocity studies and insight into supersonic turbulence, when the velocity information is not available.
\end{abstract}

\maketitle

\section{Introduction}
\label{sec:intro}

The interstellar medium (ISM) is a highly compressible turbulent, magnetized fluid, exhibiting density fluctuations on all observable scales. It has been long realized by many researchers that incompressible hydrodynamic, i.e. Kolmogorov, description is inadequate for such a medium \cite[see][for review]{elmegreen04}. Scaling relations, if they were obtained for the interstellar gas, would be very helpful for addressing many problems, including the evolution of molecular clouds and star formation.

Attempts to include effects of compressibility into the interstellar turbulence description can be dated as far back as the work by \citet{weizsacker51}. There a simple model based on a hierarchy of clouds was presented. According to this picture every large cloud consists a certain number of smaller clouds, which contain even smaller clouds. For such a model \citet{weizsacker51} proposed a relation between subsequent levels of hierarchy
\begin{equation}
 {\rho _\nu}/{\rho_{\nu - 1}} = \left( {l_\nu}/{l_{\nu - 1}} \right )^{-3\alpha} ,
 \label{eqn:hierar}
\end{equation}
where $\rho_\nu$ is the average density inside a cloud at level $\nu$, $l_\nu$ is the mean size of that cloud, 3 is the number of dimensions, and $\alpha$ is constant that reflects the degree of compression at each level $\nu$.

The Kolmogorov energy spectrum ($\sim k^{-5/3}$) follows from the assumption of a constant specific energy transfer rate $\varepsilon \sim v^2/(l/v)$. Lighthill \cite{lighthill55} pointed out that, in a compressible fluid, the volume energy transfer rate is constant in a statistical steady state
\begin{equation}
 \varepsilon_V = \rho \varepsilon \sim \rho v^2/(l/v) = \rho v^3/l.
 \label{eqn:enden}
\end{equation}

In an important, but not sufficiently appreciated work, Fleck \cite{fleck96} (henceforth, F96) incorporated above hierarchical model with energy transfer in compressible fluid to obtain the scaling relations for compressible turbulence. By combining the equations (\ref{eqn:hierar}) and (\ref{eqn:enden}) Fleck \cite{fleck96} presented the following set of scaling relations in terms of the degree of compression $\alpha$:
\begin{equation}
 \rho_l \sim l^{-3\alpha}, \ N_l \sim l^{1-3\alpha}, \ M_l \sim l^{3-3\alpha}, \ v_l \sim l^{1/3+\alpha},
 \label{eqn:relations}
\end{equation}
where $N_l$ and $M_l$ are, respectively, the column density of the fluctuation with the scale $l$ and the mass of the cloud of size $l$. The fluctuations of velocities in F96 model entail the spectrum of velocities $E(k) \sim k^{-5/3-2\alpha}$.

In the spirit of F96 model, \cite{kritsuk07} proposed to use the density-weighted velocity ${\bf u} \equiv \rho^{1/3} {\bf v}$ as a new quantity, for which the Kolmogorov scaling for second order structure functions (SFs) can be restored in compressible hydrodynamic turbulence. Their hydrodynamic simulations provided the spectrum for ${\bf u}$ close to -5/3 and they showed that in supersonic hydrodynamic turbulence the structure functions of ${\bf u}$ scale linearly with separation.

Will the F96 model be valid for compressible {\it strongly magnetized} turbulence? This is the major question that we address in this paper.

\section{Numerical Modeling}
\label{sec:numeric}

We used an second-order-accurate essentially nonoscillatory (ENO) scheme \citep[see][for details]{cho02,kowal07} to solve the ideal isothermal magnetohydrodynamic (MHD) equations in a periodic box with maintaining the $\nabla \cdot \boldsymbol{B} = 0$ constraint numerically. We drove the turbulence at wave scale $k\simeq 2.5$ (2.5 times smaller than the size of the box) using a random solenoidal large-scale driving acceleration. This scale defines the injection scale in our models. The rms velocity $\delta v$ is maintained to be approximately unity, so that $\boldsymbol{v}$ can be viewed as the velocity measured in units of the rms velocity of the system and $\boldsymbol{B}/\left(4 \pi \rho\right)^{1/2}$ as the Alfv\'{e}n velocity in the same units. The time $t$ is in units of the large eddy turnover time ($\sim L/\delta v$) and the length in units of $L$, the scale of the energy injection. The magnetic field consists of the uniform background field and a fluctuating field: $\boldsymbol{B}= \boldsymbol{B}_\mathrm{ext} + \boldsymbol{b}$. Initially $\boldsymbol{b}=0$ and $\boldsymbol{v}=0$. We use units in which the Alfv\'{e}n speed $v_A=B_\mathrm{ext}/\left(4 \pi \rho\right)^{1/2}=1$ and $\rho=1$ initially. Structures of density, velocity and magnetic field develop completely due to the forcing from uniform initial conditions. The values of $B_\mathrm{ext}$ have been chosen to be similar to those observed in the ISM turbulence. For our calculations we assumed that $B_\mathrm{ext}/\left(4 \pi \rho\right)^{1/2} \sim \delta B/\left(4 \pi \rho\right)^{1/2} \sim \delta v$. In this case, the sound speed is the controlling parameter, and basically two regimes can exist: supersonic and subsonic. Note that within our model, supersonic means low $\beta$ ($\beta \equiv p_{gas}/p_{mag}$), i.e. the magnetic pressure dominates, and subsonic means high $\beta$, i.e. the gas pressure dominates.

We present results for selected 3D numerical experiments of compressible MHD turbulence with a strong magnetic field for sonic Mach numbers ${\cal M}_s$ between $\approx 0.7$ and $7$. The Alfv\'{e}nic Mach number ${\cal M}_A\sim 0.7$. Mach numbers are defined as ${\cal M}_s \equiv \langle |v|/c_s \rangle$ and ${\cal M}_A \equiv \langle |v|/c_A \rangle$ for the sonic and Alfv\'{e}nic Mach number, respectively. To study effects of magnetization we also performed superAlfv\'{e}nic experiments with ${\cal M}_A\sim 2$. All models were calculated with the resolution 512$^3$ up to 6 dynamical times.

\section{Results}
\label{sec:resuts}

\subsection{Kolmogorov Scalings for Supersonic Flows}
\label{sec:spec_strfun}

\begin{figure}  
 \includegraphics[width=0.8\textwidth]{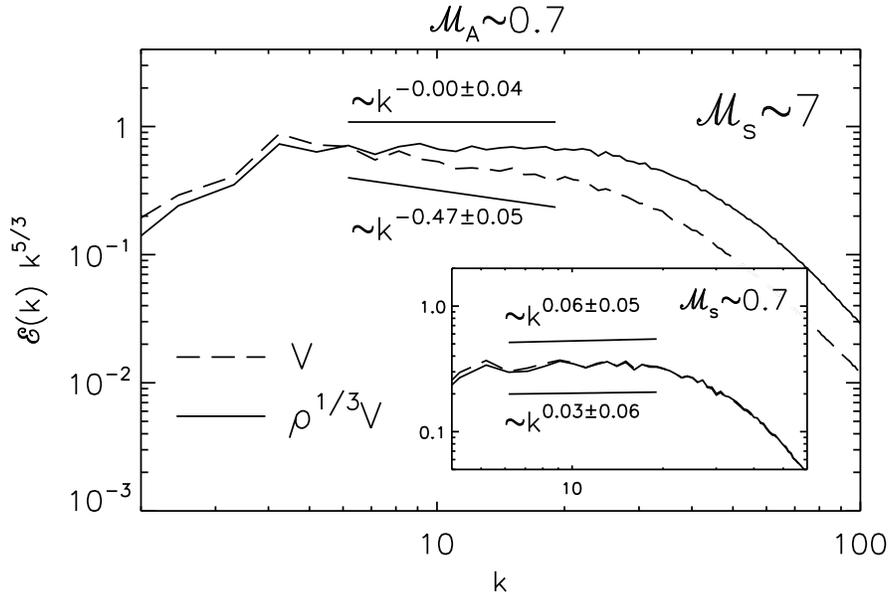}
 \caption{Spectra of velocity and density-weighted velocity (dashed and solid lines, respectively) for super- and subsonic models (big and small plots, respectively). Spectra are compensated by $k^{5/3}$. \label{fig:spectra}}
\end{figure}

In Figure~\ref{fig:spectra} we present the spectra for velocity $\boldsymbol{v}$ and density-weighted velocity $\boldsymbol{u} \equiv \rho^{1/3} \boldsymbol{v}$ for two strongly magnetized models: subsonic ($\beta \sim 2$) and supersonic ($\beta \sim 0.02$). Naturally, for subsonic model the differences between spectra for $\boldsymbol{v}$ and $\boldsymbol{u}$ are marginal and both spectra correspond to Kolmogorov's $k^{-5/3}$ scaling (see subplot in Fig.~\ref{fig:spectra}). However, we can see that for the supersonic case, the velocity spectrum gets steeper. The steepening corresponds to $\alpha\approx 0.23$ (from ${\cal E}_{v} \sim k^{-5/3-2\alpha}$). At the same time, the spectrum of $\boldsymbol{u}$ matches well the Kolmogorov slope.

\begin{figure}  
 \includegraphics[width=0.8\textwidth]{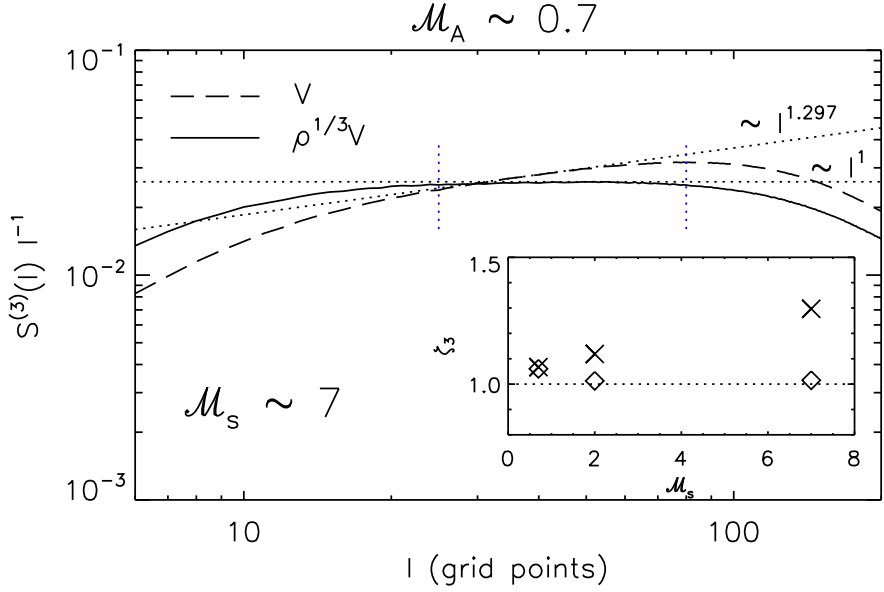}
 \caption{SFs of the third order for $\boldsymbol{v}$ (dashed line) and $\boldsymbol{u}$ (solid line) compensated by $l^{-1}$ for supersonic MHD turbulence model. Dotted lines correspond to best fitting within the inertial range. Two dotted vertical lines bound the intertial range. The sub-panel shows the scaling exponents $\zeta_3$ for $\boldsymbol{v}$ (crosses) and $\boldsymbol{u}$ (diamonds) as a function of ${\cal M}_s$. \label{fig:sf3rd}}
\end{figure}

In the original Kolmogorov theory \citep[][hereafter K41]{kolmogorov41} it was shown that the exponent of the third order structure function (SF), e.g. structure function of $\boldsymbol{v}$, $S_v^{(3)}(l) \equiv \langle | \boldsymbol{v} \left( \boldsymbol{r} + \boldsymbol{l} \right) - \boldsymbol{v}\left( \boldsymbol{r} \right) |^3 \rangle \sim l^{\zeta_3}$, should  be equal 1, i.e. $\zeta_3 = 1$. In Figure~\ref{fig:sf3rd} we show the SFs of the third order for velocity and the density-weighted velocity for supersonic model. We checked, that for the subsonic case for both $\boldsymbol{v}$ and $\boldsymbol{u}$ the index $\zeta_3$ is indeed close to unity. For the supersonic case, $\zeta_3$ increases with ${\cal M}_s$ for $\boldsymbol{v}$, but stays the unity for $\boldsymbol{u}$ (see Fig.~\ref{fig:sf3rd}). This suggests that the Kolmogorov universality is preserved for supersonic MHD turbulence when density weighting is applied.

\subsection{She-L\'{e}v\^{e}que Intermittency Model}
\label{sec:intermittency}

\begin{figure}  
 \includegraphics[width=0.8\textwidth]{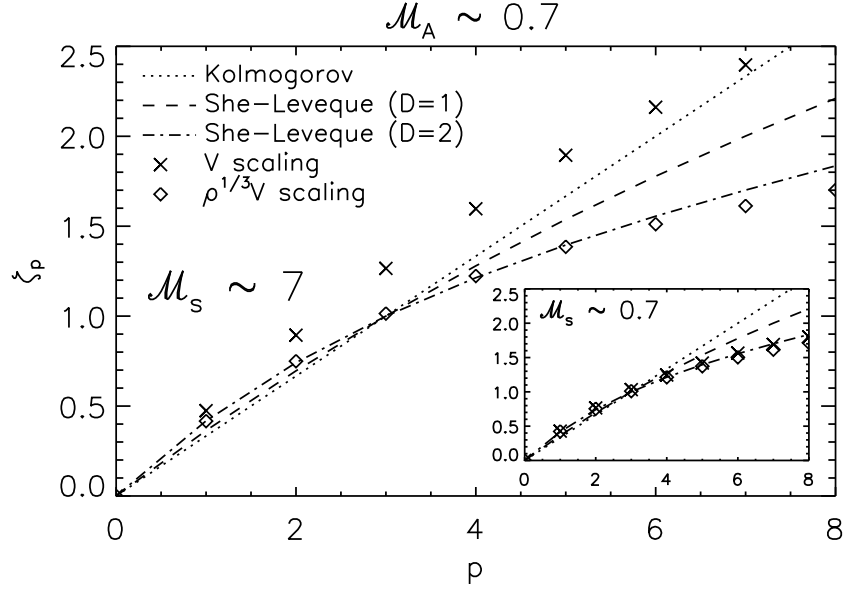}
 \caption{Scaling exponents for $\boldsymbol{v}$ and $\boldsymbol{u}$ for subsonic (subplot) and supersonic models. The plots present unnormalized values of the scaling exponents obtained directly from SFs by fitting the relation $S^{(p)}(l) = a l^{\zeta_p}$ within the intertial range, i.e. without using the extended self-similarity \citep{benzi93}. \label{fig:scalexp}}
\end{figure}

A proper description of turbulence requires higher moments \cite[see][for review]{lazarian06a}. Those characterize intermittency, which is in the original K41 model is not accounted for. A substantial progress in understanding turbulence intermittency is related to a discovery by She \& L\'{e}v\^{e}que (\cite{she94}, hereafter SL94), who found a simple form for the scaling of exponents $\zeta_p$ of higher order longitudinal correlations $S^{(p)}(l)\equiv \langle | \left[ \boldsymbol{v} \left( \boldsymbol{r} + \boldsymbol{l} \right) - \boldsymbol{v}\left( \boldsymbol{r} \right) \right] \cdot \hat{\boldsymbol{l}} |^p \rangle \sim l^{\zeta_p}$. While in K41 model $\zeta_p\equiv p/3$, SL94 provides (after modifications by M\"uller \& Biskamp \cite{mueller00} to more general form)
\begin{equation}
 \zeta_p = \frac{p}{g} (1-x) + (3-D) \left[ 1 - (1 - x/(3-D))^{p/g} \right],
\end{equation}
where $g$ is related to the scaling of the velocity $v_l \sim l^{1/g}$, $x$ is related to the energy cascade rate $t^{-1}_l \sim l^{-x}$ and $D$ is the dimension of the dissipative structures. In hydrodynamic incompressible turbulence, we have $g = 3$ and $x = 2/3$. For MHD turbulence the dissipation happens in current sheets, which are two-dimensional dissipative structures, corresponding to $D=2$ \citep{mueller00}. Thus, for subsonic MHD turbulence we expect $\zeta_p = \frac{p}{9} + 1 - (1/3)^{p/3}$ for both velocity and the density-weighted velocity. This is what we actually observe in Figure~\ref{fig:scalexp} (see subpanel). The same scaling, however, is preserved for $\boldsymbol{u}$ for {\it supersonic} magnetized turbulence.

\subsection{Anisotropies Induced by Magnetic Field}

Magnetic field is known to induce anisotropies of compressible MHD turbulence \cite[see][]{higdon84}. Anisotropy increasing with the decrease of scale was predicted for Alfv\'{e}nic motions by Goldreich \& Sridhar (\cite{goldreich95}, henceforth GS95, see also \cite{lithwick01}) and confirmed numerically for compressible MHD in \cite{cho02,cho03}.
\begin{figure}  
 \includegraphics[width=0.8\textwidth]{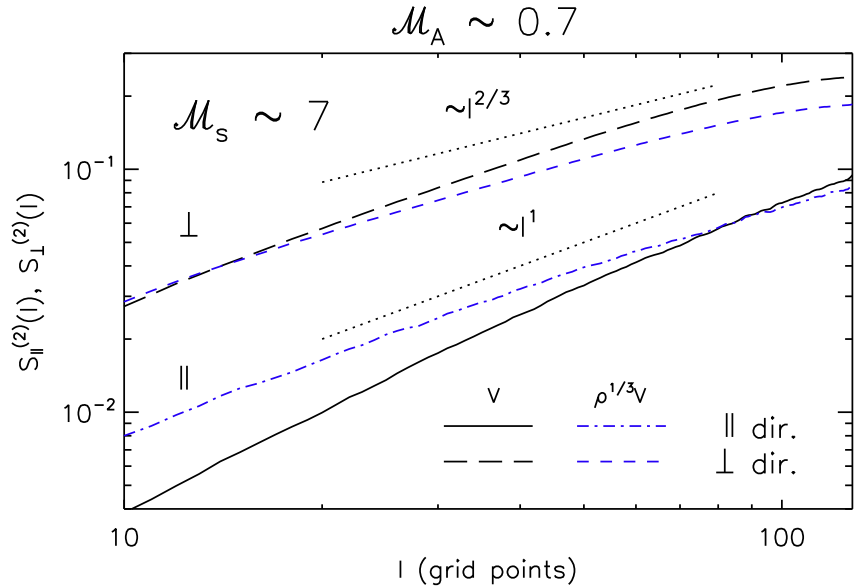}
 \caption{SFs of the second order for $\boldsymbol{v}$ and $\boldsymbol{u}$ in the local reference frame for the supersonic experiment. SFs for velocity scale as $\sim l^{1.215}$ and $\sim l^{0.874}$ for parallel and perpendicular directions to the local magnetic field, respectively. SFs for $\boldsymbol{u}$ scale as $\sim l^{0.882}$ and $\sim l^{0.744}$ for $\parallel$ and $\perp$ directions, respectively. \label{fig:local_frame}}
\end{figure}

For supersonic motions Figure~\ref{fig:local_frame} shows that the scalings for $\boldsymbol{v}$ are much steeper in both directions than those predicted by GS95 model ($1.215$ and $0.874$ for $\parallel$ and $\perp$ directions to the local magnetic field, respectively). However, those slopes still give a close to GS95 anisotropy ($l_{\|}\sim l_{\bot}^{0.718}$), which is indicative of the dominance of the Alfv\'{e}nic (``incompressible'') motions. Note, that in Figure~\ref{fig:local_frame} the SFs are obtained in the system of reference of the {\it local} magnetic field, i.e. the field on the scales of the fluctuations under study. $S^{(2)}_{\|}$ and $S^{(2)}_{\bot}$ denote second order SFs parallel and perpendicular to local magnetic field, respectively.

\begin{figure}  
 \includegraphics[width=0.5\textwidth]{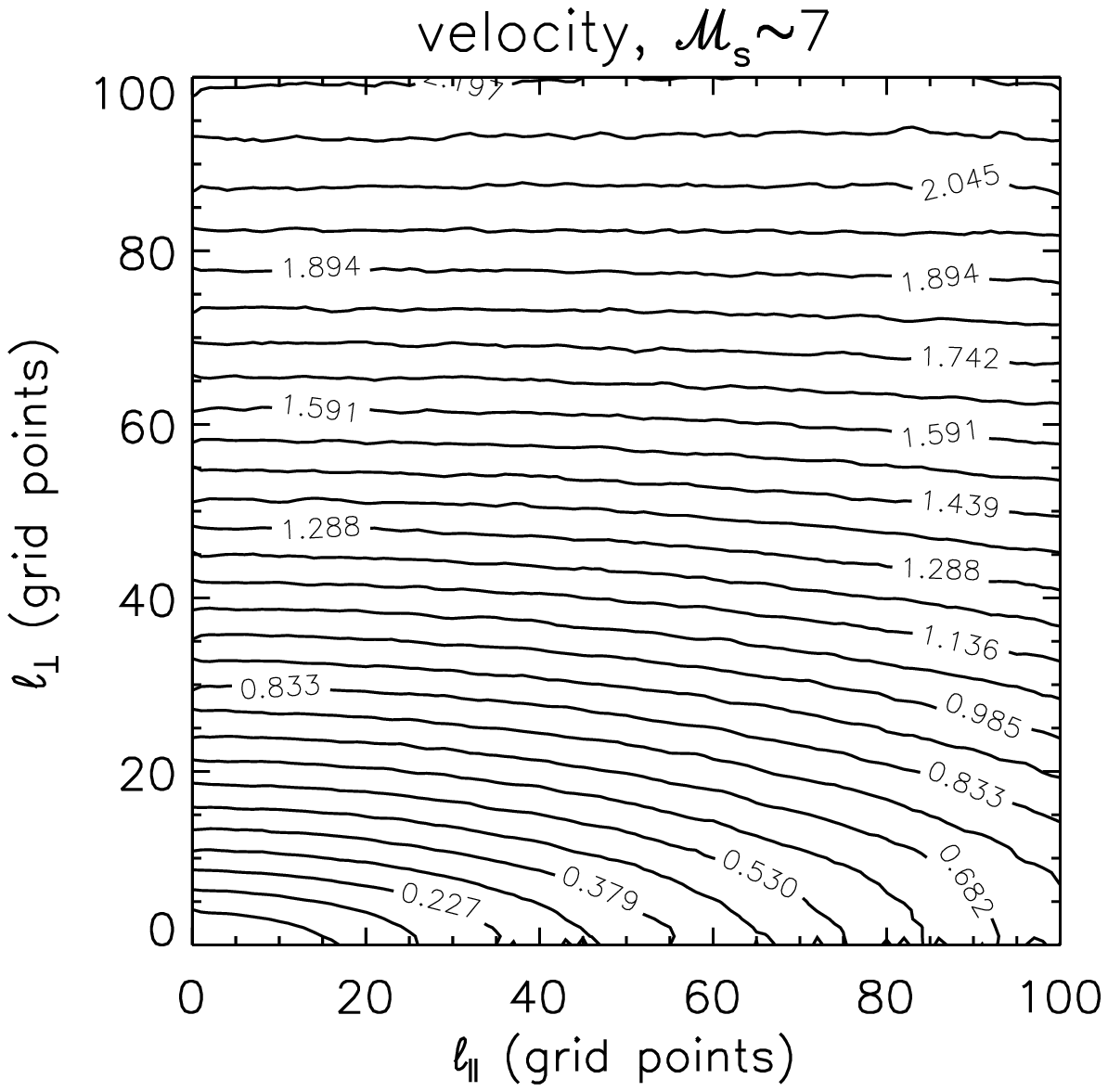}
 \includegraphics[width=0.5\textwidth]{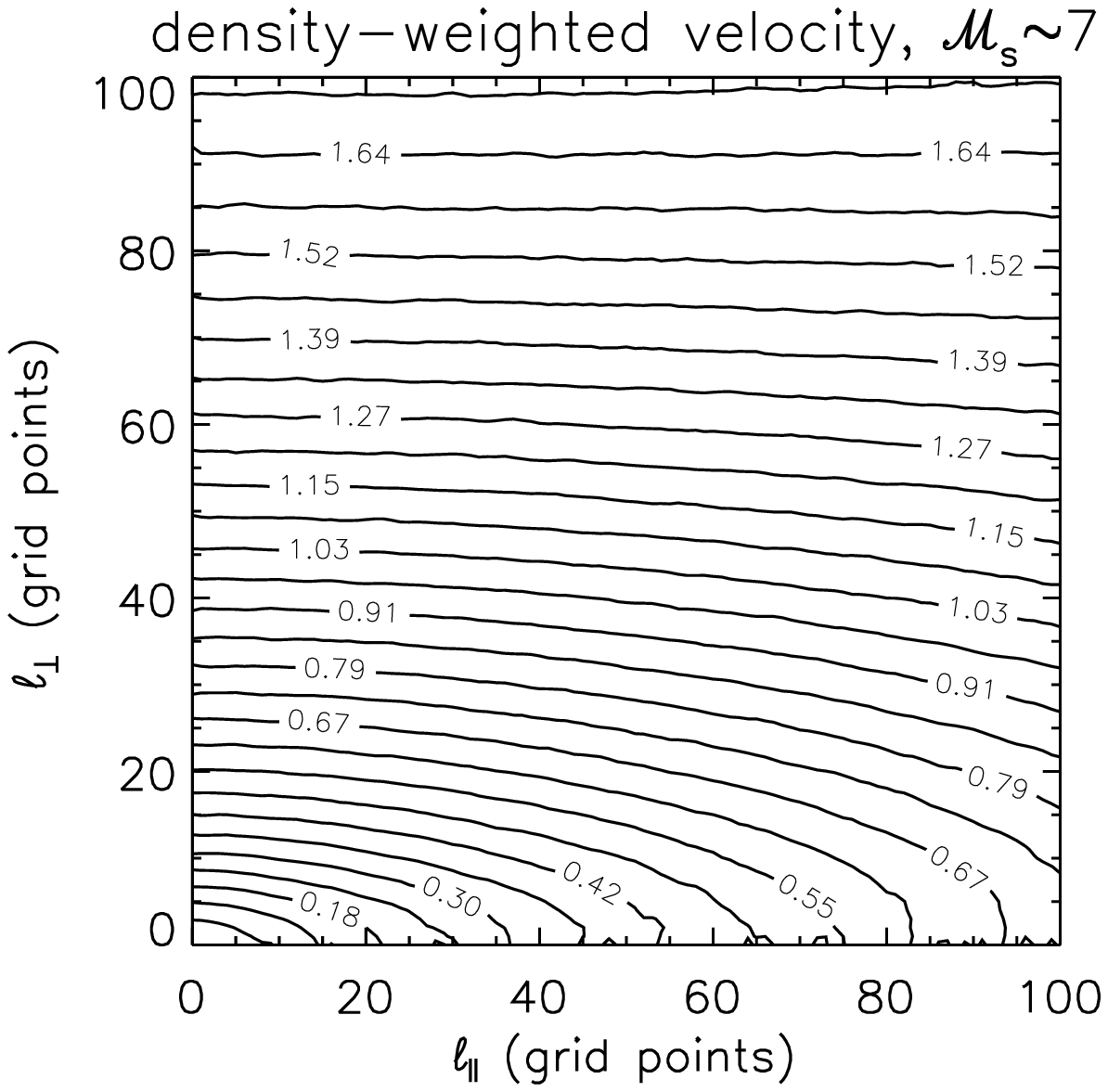}
 \caption{Contours of SFs of the second order for $\boldsymbol{v}$ and $\boldsymbol{u}$ (left and right plots, respectively) in the local reference frame for the supersonic experiment. The small-scale structures show the reduced  degree of anisotropy. \label{fig:contours}}
\end{figure}

For $\boldsymbol{u}$ the scalings are significantly shallower ($0.882$ and $0.744$ for $\parallel$ and $\perp$ directions to the local magnetic field, respectively). The SF in perpendicular direction scales more like incompressible motions, i.e. $S^{(2)}_{\bot}\sim l^{2/3}_{\bot}$. The slope of $S^{(2)}_\parallel$ for $\boldsymbol{u}$ is smaller than the corresponding one for $v$ resulting in the reduced degree of anisotropy ($l_{\|}\sim l_{\bot}^{0.843}$). This is well visible for small-scale structures as presented in the Figure~\ref{fig:contours}. Intuitively, this can be understood in terms of dense clamps strongly distorting magnetic field as they move in respect to magnetized fluid.

\subsection{Spectra for Density Fluctuations}
\label{sec`:density_spectra}

Our results for velocity show that our simulations of strongly magnetized turbulence provide $\alpha\simeq0.23$ for ${\cal M}_s \sim 7$. The spectrum of density fluctuations ${\cal E}_\rho \sim k^{-1 + 6 \alpha}$ follows from the scaling relation of density (see Eq.~\ref{eqn:relations}) according to the F96 model. This suggests the existence of a rising spectrum of density fluctuations within the hierarchy of density clumps when $\alpha > 1/6$.

\begin{figure}  
 \includegraphics[width=0.8\textwidth]{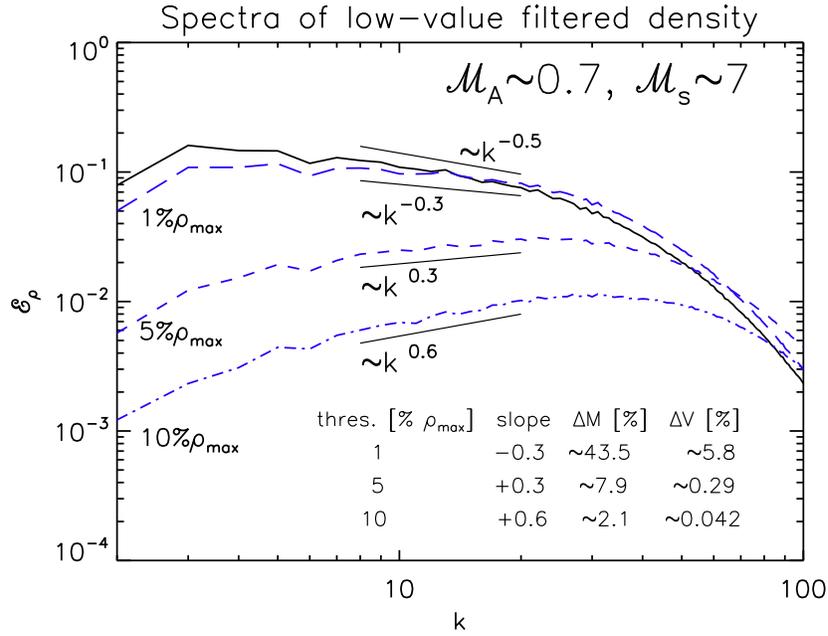}
 \caption{Spectra of the low-value filtered densities for strongly magnetized supersonic turbulence for three thresholds: 1\%, 5\% and 10\% of the maximum value of $\rho$ (long dashed, short dashed and dot-dashed lines, respectively). Solid line correspond to the spectrum of density without filtering. The embedded table shows values of spectral indices and percent of the total mass and total volume occupied by the filtered density for subsequent thresholds. \label{fig:densspec}}
\end{figure}
The Figure~\ref{fig:densspec} shows spectra of the low-value filtered densities for supersonic model. Spectra for densities above a threshold of 5\% of the maximum value start to be rising. The Figure~\ref{fig:densspec} contains also a table of mass and volume filling factors, which shows that with increasing threashold, the same amount of matter occupies less space.

\subsection{Statistics of Column Density Peaks}
\label{sec:fractal_dimension}

We try to make our study more related to {\it observations} which usually measure densities integrated along the line of sight, i.e. column densities, or alternatively study the hierarchy of observed clump masses (see Eq.~\ref{eqn:relations}). F96 model assumes the existence of an infinitely extended hierarchy. In our computations the structures are generated by turbulence at scales less than the scale of the computational box. Therefore the F96 scaling relations (in Eq.~\ref{eqn:relations}) should be modified as follows
\begin{equation}
 N_{l} \sim L\cdot l^{-3\alpha} \sim l^{-3\alpha} \mathrm{and} \ M_{l} \sim L\cdot l^{2-3\alpha} \sim l^{2-3\alpha}.
\end{equation}

Our procedure of obtaining the scaling relation from column density maps is similar to that in \cite{kritsuk07}, with the difference that they dealt with 3D data, while we deal with 2D data. First, we seek for a local maximum of column density. Then we calculate the average column density within concentric boxes with gradually increasing the size $l$. In case of determining the relation for $M_{l}$, instead of averaging we apply the integration over the boxes. Naturally, the results should correspond to each other.
\begin{figure}  
 \includegraphics[width=0.5\textwidth]{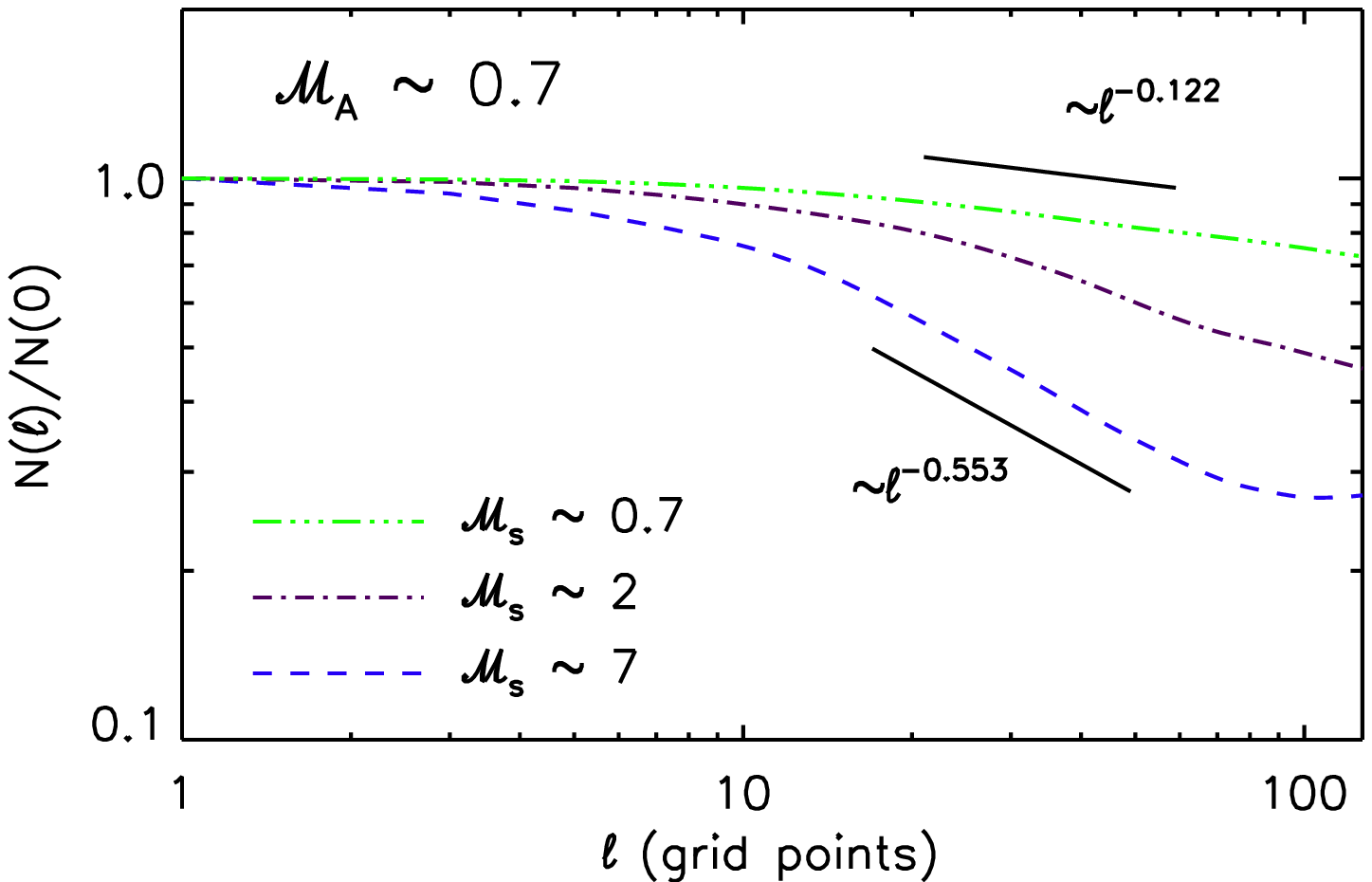}
 \includegraphics[width=0.5\textwidth]{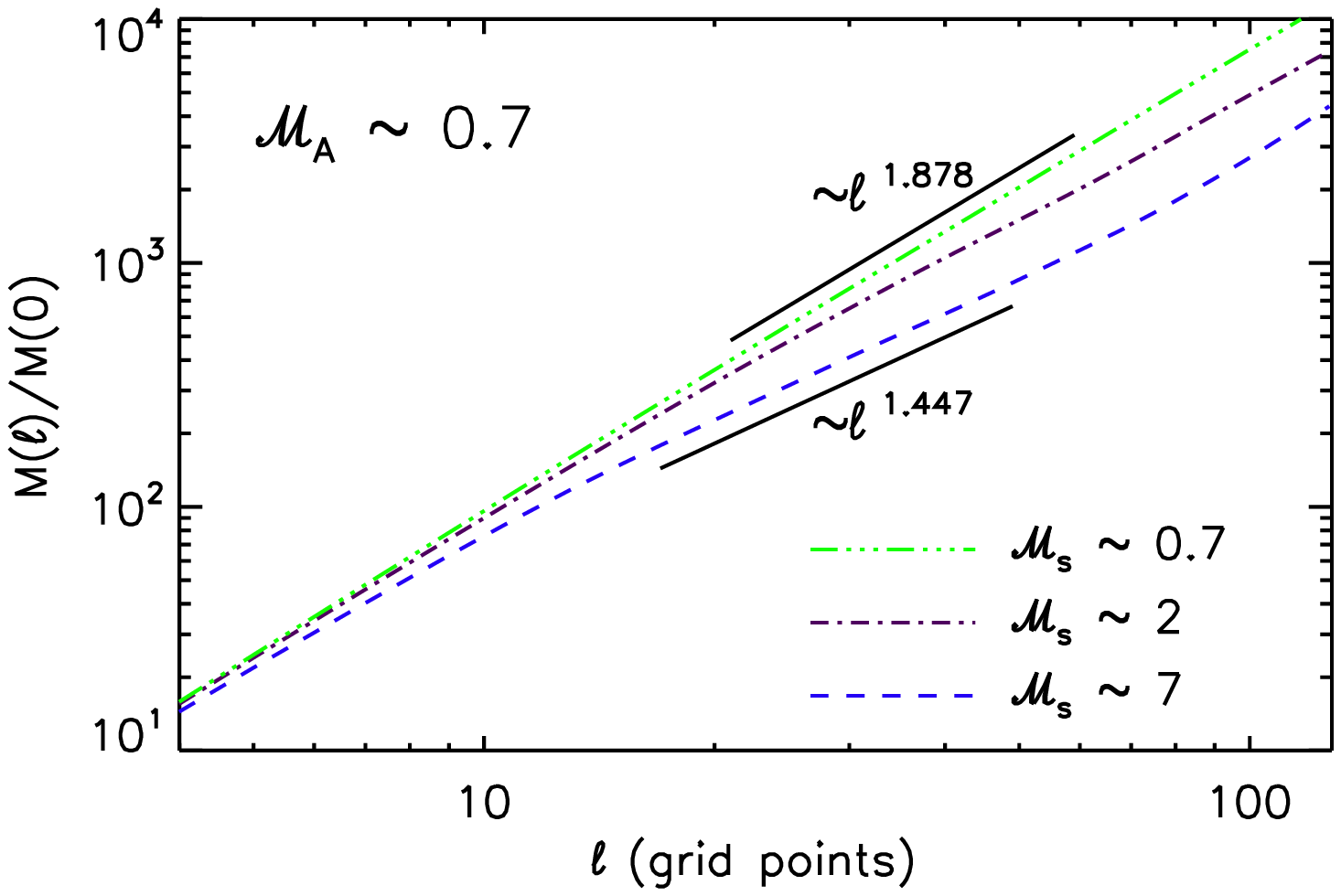}
 \caption{Scaling relations for the column density $N$ (left plot) and for density (right plot) for three models of turbulence: subsonic (${\cal M}_s\sim0.7$) and supersonic (${\cal M}_s\sim$ 2 and 7, see legend). \label{fig:densmass}}
\end{figure}

In Figure~\ref{fig:densmass} we present an example of scaling relation for column density and density for three models of turbulence with ${\cal M}_s \sim$ 0.7, 2, and 7. One can note, that the relation becomes more steep with the sonic Mach number within the intertial range. The fractal dimensions can be calculated from the relation $D_m = 3 + \gamma$ \cite[see][]{kritsuk07}, where $\gamma \equiv - 3 \alpha$ is a slope estimated from the plot within the intertial range. For our models the fractal dimension ranges from $D_m \simeq 2.5$ for the highly supersonic models to $D_m \simeq 2.9$ for subsonic model. Respectively, the compressibility coefficients for presented models $\alpha \simeq 0.04$ for ${\cal M}_s \sim 0.7$ to $\alpha \simeq 0.19$ for ${\cal M}_s \sim 7$. The latter roughly consistent with $\alpha$ obtained for the velocity SF measurement in Section ``Kolmogorov Scalings for Supersonic Flows''. The differences are probably due to insufficient statistics of rather rare high density peaks. In general, the filling factor of a peak decreases with the maximum density of this peak, which means that the higher maximum of density the peak has the smaller space it occupies.

\subsection{Variations of Scalings Induced by Fluid Magnetization}

What is the effect of magnetic field on the $\boldsymbol{u}$-scaling? The spectra, third and higher moments of correlations obtained for our superAlfv\'{e}nic simulations with ${\cal M}_A \sim 2$ happen to be very similar to the case of strongly magnetized turbulence. Our results indicate that, unlike velocity, $\boldsymbol{u}$ is much less affected by magnetic field. Naturally, in super-Alfv\'{e}nic turbulence the anisotropies induced by magnetic field are not observed at larger scales within the intertial range (cf. the last paragraph of Section ``Astrophysical Implications''.

\section{Astrophysical Implications}
\label{sec:discussion}

\paragraph{Dependence of $\alpha$ on the extend of inertial range} If we combine several facts together, namely, (a) that $\alpha$ is a function of Mach number, (b) that the maximum of density correspond to the dissipation scale, e.g. shock thickness scale $l_{diss}$, (c) that the amplitude of density in peaks scales as the mean density times ${\cal M}_s^2$, we have to conclude that as the inertial range from the injection scale $l_{inj}$ to $l_{diss}$ increases, for a given Mach number, $\alpha$ should decrease. Connecting these facts we get the following relations, $\rho_{peak} \sim {\cal M}_s^2 \sim (l_{inj}/l_{diss})^{3 \alpha}$, which gives the dependence of $\alpha$ on $l_{inj}/l_{diss}$ and ${\cal M}_s$, namely, $\alpha \sim \log{{\cal M}_s}/\log{(l_{inj}/l_{diss})}$. An interesting consequence of this would be a prediction of Kolmogorov scaling for supersonic {\it velocities} when the injection and dissipation ranges are infinitely separated. Consequently, the steeper velocity spectra reported in \cite{padoan07} can be interpreted as an indication of a limited inertial range. Further research justifying such conclusions is required, however.

\paragraph{SINS of supersonic turbulence} Ubiquitous small ionized and neutral structures (SINS) are observed in interstellar medium \cite[see][]{heiles07}. Their nature is extremely puzzling if one thinks in terms of Kolmogorov scalings for density fluctuations. The fact that the spectrum of fluctuations of density in supersonic turbulence is shallower that the Kolmogorov one is well-known \cite[see][and references therein]{kowal07}. However, just the difference in slope cannot explain the really dramatic variations in column densities observed. The present paper provides a different outlook at the problem of SINS. We see that, while low amplitude density fluctuations exhibit Kolmogorov scaling \citep{beresnyak05,kowal07}, high peaks of density correspond to a rising spectrum of fluctuations. Thus, observing supersonic turbulence at small scales, we shall most frequently observe small amplitude fluctuations corresponding to Kolmogorov-like spectrum of density fluctuations. Occasionally, but inevitably, one will encounter isolated high density peaks. An alternative mechanism for getting infrequent large density fluctuations over small scales is presented in \cite{lazarian06b} and is related to current sheets in viscosity-damped regime of MHD turbulence.

\paragraph{Clumps and star formation} Interstellar medium is known to be clumpy. Frequently clumps in molecular clouds are associated with the action of gravity. Our study shows that supersonic turbulence tend to produce small very dense clumps. If such clumps happen to get Jean's mass, they can form stars. Therefore, star formation is inevitable in supersonic turbulence. However, the efficiency of star formation is expected to be low, as the filling factor of peaks decreases with the increase of the peak height. Inhibiting of star formation via shearing may dominate in terms of influencing of star-formation efficiencies.

\section{Summary}
\label{sec:summary}

In paper above we have studied the scaling of supersonic MHD turbulence. We found that:
\begin{itemize}
 \item Fleck 1996 model is applicable to strongly magnetized compressible
turbulence.
 \item Spectra and structure functions of density-weighted velocities are
consistent with predictions of the Kolmogorov theory.
 \item Intermittency of density-weighted velocity can be well described by the
She-L\'{e}v\^{e}que model with the dimension of dissipative structures equal 2
\citep{mueller00}.
 \item Strongly magnetized supersonic turbulence demonstrate lower degree of
anisotropy if described using the density-weighted velocity.
 \item The high peaks of column densities exhibit increase of the mean values of
column densities with the decrease of scale, which may be relevant to the
explanation of SINS.
\end{itemize}

{\bf Acknowledgments}
The research of Grzegorz Kowal and Alex Lazarian is supported by the NSF grant
AST0307869 and the Center for Magnetic Self-Organization in Astrophysical and
Laboratory Plasmas. We thank Alexei G. Kritsuk for valuable discussion.


\end{document}